\documentclass[12pt]{iopart}

\expandafter\let\csname equation*\endcsname=\relax
\expandafter\let\csname endequation*\endcsname=\relax
\usepackage{amsmath}

\usepackage{bm,graphicx}
\usepackage[utf8]{inputenc}
\usepackage[T1]{fontenc}
\usepackage{lmodern}
\usepackage{amssymb}
\usepackage{bm}
\usepackage[pdftex,colorlinks,linkcolor=blue,citecolor=blue,urlcolor=blue]{hyperref}

\usepackage[normalem]{ulem}

\begin{document}
\title{Transport of ultracold atoms between concentric traps via spatial adiabatic passage}
\author{J.~Polo$^1$,
A.~Benseny$^2$,
Th.~Busch$^2$,
V.~Ahufinger$^1$, and
J.~Mompart$^1$}

\address{$^1$ Departament de F\'{\i}sica, Universitat Aut\`{o}noma de Barcelona, E-08193 Bellaterra, Spain} 
\address{$^2$ Quantum Systems Unit, Okinawa Institute of Science and Technology Graduate University, Okinawa, 904-0495, Japan}

\date{\today}

\begin{abstract}
Spatial adiabatic passage processes for ultracold atoms trapped in tunnel-coupled cylindrically symmetric concentric potentials are investigated. Specifically, we discuss the matter-wave analogue of the rapid adiabatic passage (RAP) technique for a high fidelity and robust loading of a single atom into a harmonic ring potential from a harmonic trap, and for its transport between two concentric rings. We also consider a system of three concentric rings and investigate the transport of a single atom between the innermost and the outermost rings making use of the matter-wave analogue of the stimulated Raman adiabatic passage (STIRAP) technique. We describe the RAP-like and STIRAP-like dynamics by means of a two- and a three-state models, respectively, obtaining good agreement with the numerical simulations of the corresponding two-dimensional Schr\"odinger equation.
\end{abstract}

\maketitle

\section{Introduction}
\label{sec:introduction}

Toroidal trapping potentials \cite{rings,rings_2,rings_3,rings_4,rings_5,rings_6} for ultracold atoms are essential in many quantum metrology implementations, e.g., for matter-wave Sagnac interferometry \cite{Sagnac}, and as elementary building blocks in the emerging field of atomtronics \cite{atomtronics}. In fact, ring potentials constitute the simplest nontrivial closed-loop circuits and offer unique features to investigate, for instance, superfluidity and persistent currents \cite{persistentcurrents}.
Angular momentum can be transferred to the atoms in such systems using either light \cite{persistentcurrents_experiment1} or by rotating a tunable weak link \cite{persistentcurrents_experiment2}.
The latter has been one of the key elements that has opened the possibility to study matter-wave analogues of the superconducting quantum interference devices (SQUIDs) \cite{new_atomtronics}.
Recently, hysteresis, which is a basic ingredient in electronics, has been observed experimentally between quantized circulation states of a Bose--Einstein condensate (BEC) trapped in a ring \cite{rings_1}.

BECs in two tunnel-coupled concentric ring potentials have also been investigated, paying special attention to the generation of topological phases under rotation \cite{Brand_2009}. Ground-state phases, produced by the interplay between two-body collisions and spin-orbit coupling \cite{Zhang_2012} or between the intra-atomic and inter- atomic interactions in a two-component BEC \cite{Malet_2010}, have also been analyzed.
In this context, the development of techniques allowing for a high fidelity and robust transport of ultracold atoms between ring potentials is a topical issue.
Spatial adiabatic passage (SAP) processes \cite{review_SAP} have been proposed for the transport of single atoms 
\cite{eckert1,eckert2,singleatom_1DSAP,loiko,morgan}, electrons \cite{electrons_1DSAP}, and Bose--Einstein condensates \cite{BEC_1DSAP} between the outermost traps of in-line triple-well potentials. Moreover, two-dimensional (2D) SAP processes for three tunnel-coupled harmonic wells in a triangular configuration have recently been discussed for matter-wave interferometry \cite{interferometry_2DSAP}, and for the generation of atomic states carrying orbital angular momentum \cite{OAM_2DSAP}. It is also worth to highlight that SAP has already been experimentally demonstrated for light beams propagating in three evanescently coupled optical waveguides \cite{light_SAP}, and  recently discussed for the manipulation of sound propagation in sonic crystals \cite{sound_SAP}.

In this work, we investigate the use of 2D SAP processes for the transport of ultracold atoms between the localized ground states of cylindrically symmetric concentric potentials. Specifically, we focus on the matter-wave analogues of the well-known quantum-optical rapid adiabatic passage (RAP) \cite{RAP} and stimulated Raman adiabatic passage (STIRAP) \cite{STIRAP} techniques.
We consider the matter-wave analogue of the RAP technique for the loading of a single atom from a harmonic trap into a ring potential, see Fig.~\ref{fig:1POT}(a), or for transporting it between two concentric ring potentials, see Fig.~\ref{fig:1POT}(b). For these RAP-like processes, we vary the trapping frequency and the radius of the outer ring potential in order to allow for adjustments of the tunneling rate and the energy bias in such a way that the atom adiabatically follows a particular energy eigenstate of the system. Moreover, by means of a STIRAP-like technique, we investigate the transport of a single atom between the innermost and outermost traps of a triple-ring potential, see Fig.~\ref{fig:1POT}(c). In this case, we assume equal radial trapping frequencies for the three harmonic rings and the tunneling-based transport is achieved by modifying the rings separations. Both SAP processes studied here rely on the ability to vary the frequencies and radii of the trapping potentials.
For optically generated potentials, the frequency can be modified by varying the laser intensity that produces the potential \cite{rings_2}, and the radius of the ring can be changed dynamically using time-averaged adiabatic techniques \cite{rings_3,rings_4}.
These matter-wave analogues of RAP and STIRAP are described with a two- and a three-state model, respectively, and the obtained results are found to be in good agreement with the corresponding numerical simulations of the 2D Schr\"odinger equation. Finally, we also investigate the possibility to use SAP processes between three concentric ring potentials to transport orbital angular momentum (OAM) states.
Such transport can play a very important role in future atomtronic devices, where OAM states are one of the building blocks to study SQUIDs \cite{new_atomtronics}.
However, angular momentum states also introduce extra degrees of freedom which  consequently will affect the adiabaticity condition of the 2D SAP processes.
It is therefore necessary to check if the adiabatic techniques are still usable in ring type geometries.
For instance,
in 1D systems, the energy biases are only affected by the difference in frequencies of the traps.
In 2D SAP systems, however,
the energies of the rings also depend on their radii and thus, coupling and detuning cannot be treated as independent parameters.

The article is structured as follows.
The physical system under investigation is introduced in Section~\ref{sec:physicalsystem}. Section~\ref{sec:two_potentials} is devoted to the transport of a single atom between two concentric potentials by means of the matter-wave RAP technique, whereas Section~\ref{sec:three_potentials} focuses on the transport of an atom between the innermost and outermost traps of a triple-ring potential through the matter-wave STIRAP technique. Section~\ref{sec:conclusions} presents the conclusions.

\section{Physical system}
\label{sec:physicalsystem}

\begin{figure}
\centerline{\includegraphics[width=0.85\textwidth]{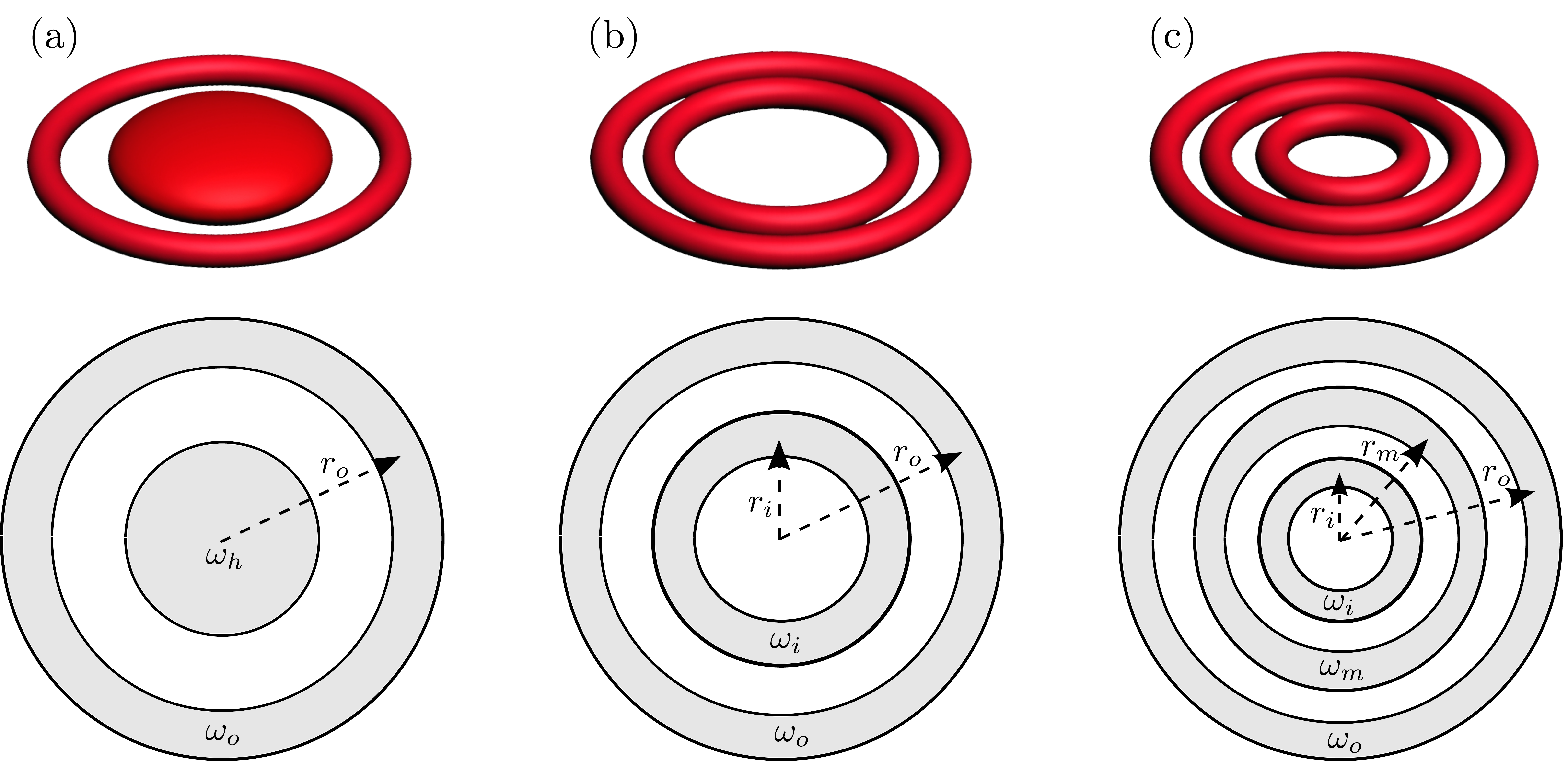}}
	\caption{Sketch of the trapping geometries formed by the combination of concentric (a) harmonic and harmonic ring potentials; and (b) two and (c)  three harmonic ring potentials. Transport will always take place from the innermost trap to the outermost one.
	The top row shows three-dimensional representations of the trapping potentials. 
	The bottom row shows 2D schematics of the three configurations, with traps represented in gray, and including labels for trap frequencies and radii.
	$\omega_h$ is the radial trapping frequency of the harmonic potential; $r_j$ and $\omega_j$ with $j=i,m,o$ are the radii and radial trapping frequencies of the inner, middle, and outer rings, respectively. 
		}
	\label{fig:1POT}
\end{figure}

We consider a single atom trapped in 2D cylindrically symmetric external potentials formed by different concentric combinations of harmonic and  harmonic ring potentials, see Fig.~\ref{fig:1POT}.
Using cylindrical coordinates $(r,\phi)$, the harmonic potential with trapping frequency $\omega_h$ for an atom of mass $m$ reads
\begin{equation}
\label{eq:hp}
	V_h(r)=\frac{1}{2}m\omega_h^{2}r^{2},
\end{equation}
and its ground state wavefunction is
\begin{equation}
\label{eq:grdh}
	\Psi_h(r)=\sqrt{\frac{m\omega_h}{\pi\hbar}}e^{-\frac{m\omega_h}{2\hbar}r^{2}}.
\end{equation}

The harmonic ring potentials of trapping frequencies $\omega_j$ and radii $r_j$ are given by  
\begin{equation}
\label{eq:hr}
	V_{r,j}(r)=\frac{1}{2}m\omega_j^2(r-r_j)^2,
\end{equation}
with $j=i,m,o$ labeling the inner, middle and outer rings, respectively, as shown in Fig.~\ref{fig:1POT}. 
In order to obtain the lowest energy eigenstate of a single atom in a harmonic ring potential, we use a variational approach with the ansatz
\begin{equation}
\label{eq:grdr}
	\Psi_{r,j}(r)=N_j e^{-\alpha_j (r-\beta_j r_j)^2},
\end{equation}
where $\alpha_j$ and $\beta_j$ are the variational parameters, and $N_j$ is fixed by the normalization condition $2\pi\int^{\infty}_{0}|\Psi_{r,j}(r)|^{2} r dr=1$, which leads to
\begin{equation}
	N_j = \sqrt{\frac{2 \alpha_j/\pi }{e^{-2 \alpha_j  \beta_j ^2 r_j^2}+ \beta_j  r_j\sqrt{2 \pi \alpha_j} \left(\textrm{erf}\left(\beta_j  r_j\sqrt{2\alpha_j} \right)+1\right)}}.
\end{equation}

The joint potentials $V_T(r)$ sketched in Fig.~\ref{fig:1POT} are modeled as piecewise combinations of the corresponding potentials given by (\ref{eq:hp}) and (\ref{eq:hr}), and are truncated at the points where they coincide (see Eqs.~(\ref{eq:vrap1}), (\ref{eq:vrap2}) and (\ref{eq:vrap3})).
Truncated harmonic potentials have shown to give qualitatively similar results for SAP as smooth Gaussian \cite{eckert2}, P\"oschl--Teller potentials \cite{loiko}, or atom-chip systems \cite{morgan}.
Moreover, they allow to use a simple ansatz for the localized ground states such as the ones given in Eqs.~(\ref{eq:grdh}) and (\ref{eq:grdr}).

The dynamics of a single atom in this potential are then governed by the 2D Schr\"odinger equation
\begin{equation}
i\hbar\frac{\partial}{\partial t}\Psi(r,\phi,t)=H\Psi(r,\phi,t),
\label{eq:schr}
\end{equation}
where the Hamiltonian is
\begin{equation}
H =
-\frac{\hbar^{2}}{2m}
\left( 
\frac{\partial^{2}}{\partial r^{2}}
+\frac{1}{r}\frac{\partial}{\partial r}
+\frac{1}{r^2}\frac{\partial^{2}}{\partial \phi^{2}}
\right)
+V_T(r).
\end{equation}
Note that the azimuthal dependence $\phi$ in (\ref{eq:schr}) does not play any role in the dynamics if they are restricted to the ground states of the harmonic and harmonic ring potentials for which $\Psi(r,\phi,t)=\Psi(r,t)$. In the following, we will focus mainly on these states, although states carrying OAM (and thus having an azimuthal phase dependence) will be addressed at the end of Section~\ref{sec:three_potentials}.

Throughout the paper, we will express all variables in dimensionless harmonic oscillator (h.o) units with respect to $V_h$ ($m=\hbar=\omega_h=1$).

\section{Spatial Adiabatic Passage between two concentric potentials}
\label{sec:two_potentials}

\subsection{Two-state model}

In this section we introduce the two-state model for the trapping configurations shown in Figs.~\ref{fig:1POT}(a) and (b). Let us assume that at any time the total atomic wavefunction can be written as a superposition of the two orthonormalized localized states of each individual potential as
\begin{equation}
\label{eq:psiTS}
	\Psi_{2S}(r,t)=a_i(t) \widetilde{\Psi}_i(r)+a_o(t) \widetilde{\Psi}_o(r),
\end{equation}
where $a_i$ ($a_o$) is the probability amplitude for the atom to be in the inner (outer) potential.

Introducing the ansatz $\Psi_{2S}$ into the Schr\"odinger equation (\ref{eq:schr}), one obtains the two-state model, 
\begin{equation}
\label{eq:matrix}
	 i\hbar
	 \frac{d}{dt}
	 \left(\begin{array}{c}
		 a_i\\
		 a_o
	 \end{array} \right) 	 
	 =\hbar
 \left(\begin{array}{cc}
  0 & -J/2 \\
  -J/2 & \Delta 
 \end{array} \right)
 	 \left(\begin{array}{c}
		 a_i\\
		 a_o
	 \end{array} \right),
\end{equation}
where
\begin{subequations}
\label{definitionsJD}
	\begin{eqnarray}
		J=&\frac{4\pi}{\hbar} \int_{0}^{\infty}{\widetilde{\Psi}_i^{*}(r)H\widetilde{\Psi}_o(r)\,rdr},\label{eq:tunneling}\\
		\Delta=&\frac{2\pi}{\hbar} \left(\int_{0}^{\infty}{\widetilde{\Psi}_o^{*}(r)H\widetilde{\Psi}_o(r)\,r dr}-\int_{0}^{\infty}{\widetilde{\Psi}_i^{*}(r)H\widetilde{\Psi}_i(r)\,r dr}\right),\label{eq:energybias}
	\end{eqnarray}
\end{subequations}
are the tunneling rate and energy bias, respectively.
The values of $J$ and $\Delta$ used in the two-state model simulations are obtained by performing these integrals numerically with the corresponding wavefunctions.
The states $\widetilde{\Psi}_j$ for $j=i,o$ are obtained by orthonormalizing the states $\Psi_h$ and $\Psi_{r,o}$ for the configuration in Fig.~\ref{fig:1POT}(a) or $\Psi_{r,i}$ and $\Psi_{r,o}$ for the configuration in Fig.~\ref{fig:1POT}(b).

The eigenvalues of the standard two-state Hamiltonian in (\ref{eq:matrix}) are
 \begin{equation}
 \label{eq:enertsm}
	 E_{\pm}=\frac{\hbar}{2}\left(\Delta\pm\sqrt{\Delta^{2}+J^{2}}\right),
 \end{equation}
with corresponding eigenstates
\begin{subequations}
  \begin{eqnarray}
	  \Psi_{+}(\theta)=&\sin(\theta)\widetilde{\Psi}_i-\cos(\theta)\widetilde{\Psi}_o, \label{estat1} \\
		\Psi_{-}(\theta)=&\cos(\theta)\widetilde{\Psi}_i+\sin(\theta)\widetilde{\Psi}_o, \label{estat0} 
	\end{eqnarray}  
\end{subequations}	
where the mixing angle $\theta$ is given by
  \begin{equation}
 \tan(2\theta)=\frac{J}{\Delta}.
 \label{eq:angle2S}
  \end{equation}

\subsection{RAP-like protocol}

By adiabatically following either $\Psi_+$ or $\Psi_-$, a single atom initially loaded into the localized ground state of the inner potential can be efficiently and robustly transferred to the outer one. Let us assume that initially, at $t=t_0$, one has $\Delta <0$ with $\left|\Delta\right| \gg J$, such that the mixing angle in (\ref{eq:angle2S}) is given by $\theta (t_0) = \pi /2$ and, according to (\ref{estat1}), $\Psi_{2S}(t_0)=\Psi_+(t_0)=\widetilde{\Psi}_i$. To transfer the atom to the outer potential by adiabatically following $\Psi_+$, one needs to slowly modify the energy bias and the tunneling rate such that at the end of the process, at $t=t_f$, $\Delta >0$ and $\left|\Delta\right| \gg J$, which, in turn, means $\theta (t_f)=0$ and $\Psi_{2S}(t_f)=\Psi_+(t_f)=-\widetilde{\Psi}_o$.
To avoid diabatic transitions and keep the atom in the energy eigenstate $\Psi_+$ for the whole dynamics, it is important to reach significant tunneling rates when the sign of the energy bias is reversed such that even at this moment the energies of the two eigenstates are substantially different, see (\ref{eq:enertsm}).

\begin{figure}
\centerline{  \includegraphics[width=0.95\textwidth]{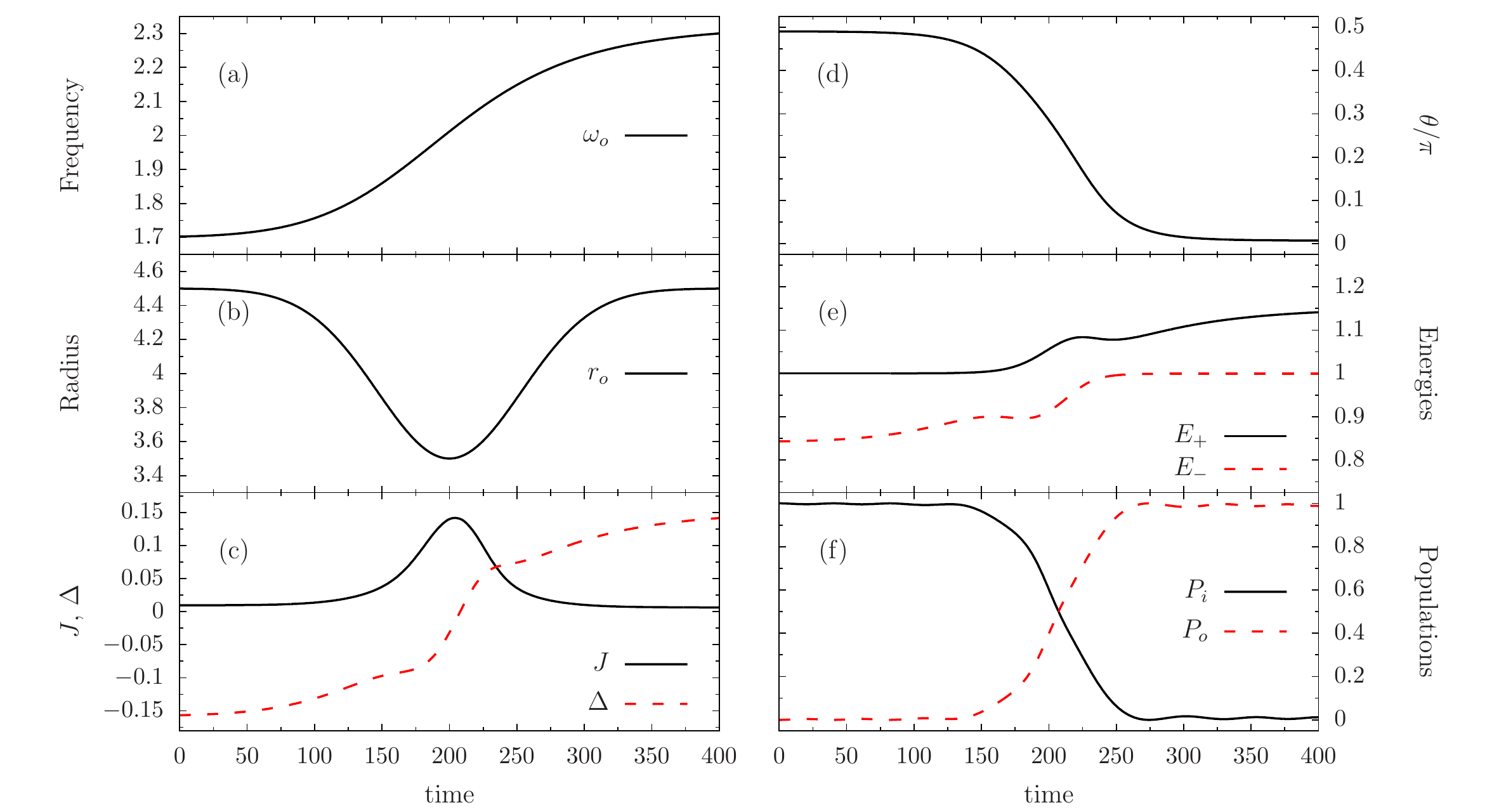}}
	\caption{Numerical simulations using the two-state model. Matter-wave RAP process to transport a single atom from the localized ground state of a harmonic potential to the ground state a harmonic ring one, see Fig.~\ref{fig:1POT}(a). As a function of time: (a) trapping frequency  $\omega_o$ and (b) radius of the harmonic ring potential $r_o$; (c) tunneling rate $J$ (black solid line) and energy bias $\Delta$ (red dashed line); (d) mixing angle $\theta$; (e) eigenvalues $E_\pm$ of the energy eigenstates $\Psi_+$ (black solid line) and $\Psi_-$ (red dashed line);  and (f) populations $P_i=|a_i|^2$ (black solid line) and $P_o=|a_o|^2$ (red dashed line) of the harmonic and harmonic ring potentials. All variables are expressed in h.o. units with respect to the potential $V_h$.}
	\label{fig:2RAP}
\end{figure}

Let us consider the transfer of a single atom trapped in a harmonic potential into a harmonic ring, see Fig.~\ref{fig:1POT}(a). In this case, the joint potential profile reads
\begin{equation}
\label{eq:vrap1}
	V_{T}(r)=
	\left\{
  \begin{array}{ll}
    V_h(r)   \quad& \textrm{for} \ r < \bar{r}_{hr}, \\
    V_{r,o}(r) \quad& \textrm{for} \ r \ge \bar{r}_{hr},
  \end{array}
\right.
\end{equation}
with $\bar{r}_{hr}=r_o \omega_o/(\omega_h+\omega_o)$.
To investigate the SAP process we modify the ring trapping frequency $\omega_o$ and radius $r_o$ according to (see Figs.~\ref{fig:2RAP}(a) and (b))
\begin{subequations}
\label{eq:woro}
\begin{eqnarray}
\label{eq:worap1}
\omega_o(t)=&\omega_0+\delta_\omega\frac{1+\left(\frac{2}{3}\right)^a}{1+\left(\frac{t}{t_f}+\frac{1}{2}\right)^{-a}} , \\
\label{eq:rorap1}
r_o(t)=& R_0 + \delta_R e^{-\frac{\left(\frac{t}{t_f}-\frac{1}{2}\right)^2}{2\sigma^2}},
\end{eqnarray}
\end{subequations}
with the parameter values $t_f=400$,
$R_0=4.5$, $\delta_R=-1$,
$\omega_0=1.7$, $\delta_\omega=0.6$, $a=8$, and $\sigma=2/15$.
These modifications of the potential yield the temporal variations of the energy bias and tunneling shown in Fig.~\ref{fig:2RAP}(c), calculated with Eqs. (\ref{definitionsJD}).
 
 \begin{figure}
\centerline{\includegraphics[width=0.75\textwidth]{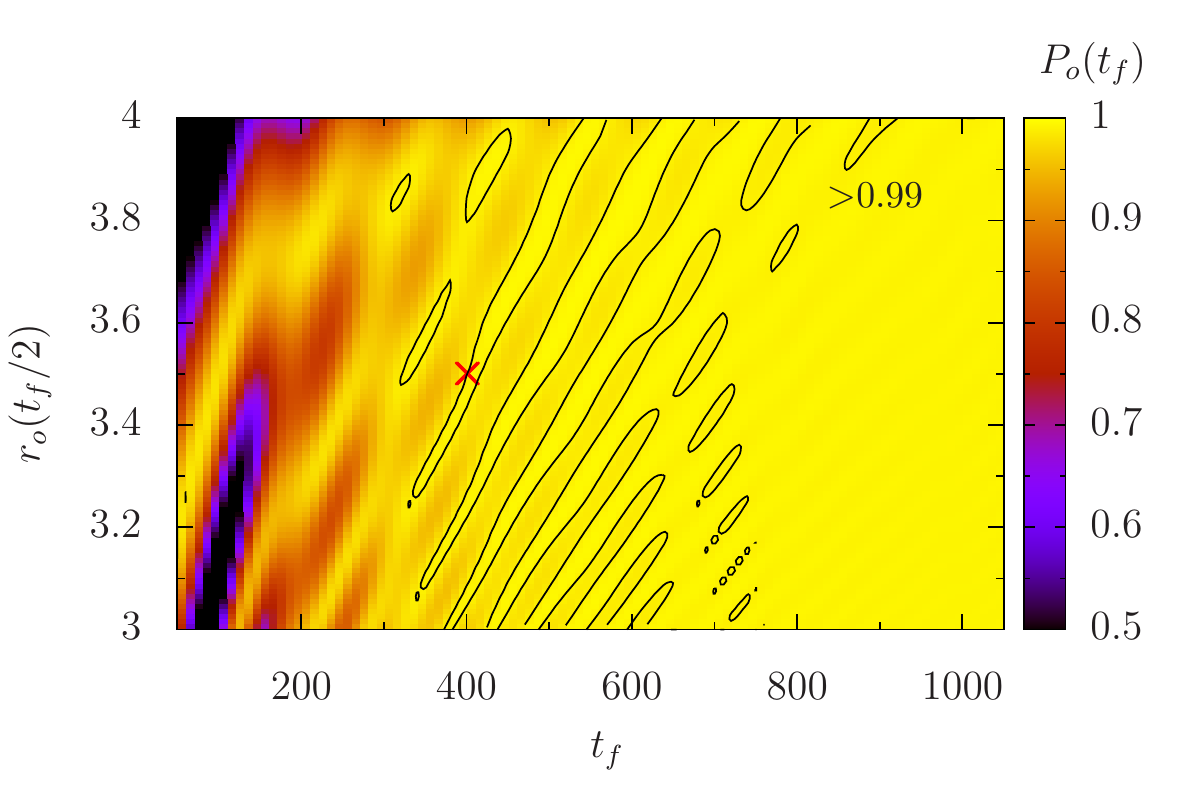}}
	\caption{Numerical integration of the 2D Schr\"odinger equation. Final population transferred to the harmonic ring trap from the inner harmonic trap as a function of the total time of the process and the minimum radius of the ring trap during the process. All other parameters are as in Fig.~\ref{fig:2RAP}.
	Black lines correspond to contours at $P_o(t_f) = 0.99$ and the cross indicates the parameter values that were chosen in the two-state simulations shown in Fig.~\ref{fig:2RAP}.
	All variables are expressed in h.o. units with respect to the harmonic potential $V_h$.
	 }
	\label{fig:3Robustness}
\end{figure}
 
The temporal evolution of the mixing angle is plotted in Fig.~\ref{fig:2RAP}(d), while Fig.~\ref{fig:2RAP}(e) shows the evolution of the energy eigenvalues (\ref{eq:enertsm}), revealing the existence of an energy gap during the whole dynamics. The populations of the harmonic and harmonic ring potentials are shown in Fig.~\ref{fig:2RAP}(f), which demonstrate the high fidelity of the transport protocol. 
Let us comment here that, as in all adiabatic protocols, the exact time dependence of the parameters is not crucial for the technique to succeed, and even withstands some noise \cite{eckert1,eckert2}.
We have chosen this particular time dependence as it allows for shorter total times $t_f$ of the protocol with respect to other simpler temporal variations of the parameters.

To check the validity of the two-state model we have also performed numerical simulations of the RAP-like protocol using the 2D Schr\"odinger equation (\ref{eq:schr}) for the same parameters as in Fig.~\ref{fig:2RAP}.
Fig.~\ref{fig:3Robustness} shows the final population transferred from the harmonic inner potential to the outer ring, $P_o$, as a function of the total time of the process and the minimum radius of the ring during the process. A population transfer above $99\%$ is achieved for a broad range of parameter values evidencing the fidelity and robustness of the process. 
The parameters corresponding to the transport protocol investigated within the two-state model above are marked with a cross in Fig.~\ref{fig:3Robustness}. 

\begin{figure}
\centerline{  \includegraphics[width=0.95\textwidth]{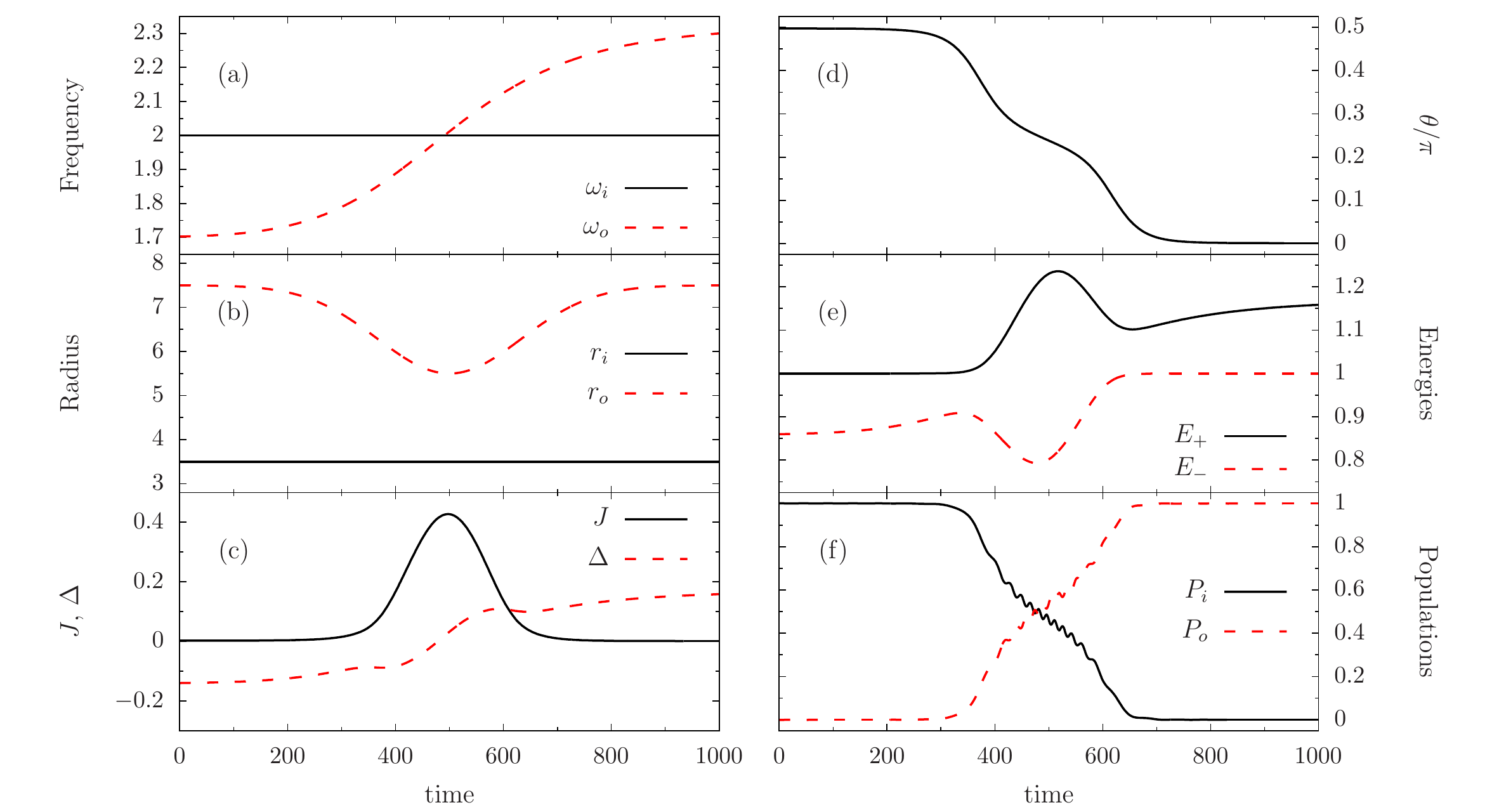}}
	\caption{Numerical simulations using the two-state model. Matter-wave RAP process to transport a single atom between two concentric ring potentials, see Fig.~\ref{fig:1POT}(b). As a function of time: (a) trapping frequencies $\omega_i$ (black solid line) and $\omega_o$ (red dashed line) of the inner and outer rings; (b) radii $r_i$ (black solid line) and $r_o$ (red dashed line) of the inner and outer rings;
	(c) tunneling rate $J$ (black solid line) and energy bias $\Delta$ (red dashed line); (d) mixing angle $\theta$; (e) eigenvalues $E_\pm$ of the energy eigenstates $\Psi_+$ (black solid line) and $\Psi_-$ (red dashed line); (f) populations $P_i=|a_i|^2$ (black solid line) and $P_o=|a_o|^2$ (red dashed line) of the inner and outer rings. All variables are expressed in h.o. units with respect to the harmonic potential $V_h$.}
	\label{fig:4RAP_rings}
\end{figure}

The previously described RAP-like technique can also be applied to the configuration shown in Fig.~\ref{fig:1POT}(b) corresponding to two concentric harmonic ring potentials described by the potential
\begin{equation}
\label{eq:vrap2}
	V_T(r)=
	\left\{
  \begin{array}{ll}
    V_{r,i}(r) \quad& \textrm{for} \ r < \bar{r}_{io} \\
    V_{r,o}(r) \quad& \textrm{for} \ r \ge \bar{r}_{io}
  \end{array}
\right.
\end{equation}
with $\bar{r}_{io} = (r_i \omega_i + r_o \omega_o)/(\omega_i + \omega_o)$.
As shown in Figs.~\ref{fig:4RAP_rings}(a) and (b), we change the frequency and radius of the outer ring in time following Eqs.~(\ref{eq:woro}) with parameters $\omega_0 = 1.7$, $\delta_\omega = 0.6$, $R_0 = 7.5$, and $\delta_R = 2$, while leaving the inner ring fixed with radius $r_i=3.5$ and frequency $\omega_i=2$.
Then, the temporal profiles of the energy bias and the tunneling rate, see Fig.~\ref{fig:4RAP_rings}(c), are such that the mixing angle $\theta$ evolves from $\pi /2$ to $0$, see Fig.~\ref{fig:4RAP_rings}(d).
In this case, the followed eigenstate $\Psi_+$ is transformed from initially being the inner ring ground state to finally being the outer ring one. Therefore, the single atom is transported from the inner ring to the outer one with very high fidelity, see Fig.~\ref{fig:4RAP_rings}(f).

\section{Spatial Adiabatic Passage between three concentric potentials}
\label{sec:three_potentials}

\subsection{Three-state model}

We now consider the trapping configuration depicted in Fig.~\ref{fig:1POT}(c) consisting of three concentric ring potentials, constructed from truncated harmonic potentials
\begin{equation}
\label{eq:vrap3}
	V_T(r)=
	\left\{
  \begin{array}{ll}
    V_{r,i}(r) \quad& \textrm{for} \ r \le \bar{r}_{im} \\
    V_{r,m}(r) \quad& \textrm{for} \ \bar{r}_{im} < r < \bar{r}_{mo} \\
    V_{r,o}(r) \quad& \textrm{for} \ r \ge \bar{r}_{mo}
  \end{array}
\right.
\end{equation}
with
$\bar{r}_{im} = (r_i \omega_i + r_m \omega_m)/(\omega_i + \omega_m)$ and
$\bar{r}_{mo} = (r_m \omega_m + r_o \omega_o)/(\omega_m + \omega_o)$.

In analogy with the previous section, we assume that the total atomic wavefunction can be written at all times as a superposition of the three orthonormalized localized ground states of each individual potential as
\begin{equation}
\label{eq:psiTS}
	\Psi_{3S}(r,t)=a_i(t) \widetilde{\Psi}_i(r)+a_m(t) \widetilde{\Psi}_m(r)+a_o(t) \widetilde{\Psi}_o(r).
\end{equation}
The equations of motion for the probability amplitudes $a_i$, $a_m$, $a_o$ of the inner, middle, and outer ring potentials, respectively, read
\begin{equation}
\label{eq:matrix2}
	 i\hbar
	 \frac{d}{dt}
	 \left(\begin{array}{c}
		 a_i\\
		 a_m \\
		 a_o
	 \end{array} \right) 	 
	 =\hbar
 \left(\begin{array}{ccc}
  \Delta_i & -J_{im}/2 & 0\\
  -J_{im}/2& 0 & -J_{mo}/2\\
  0 & -J_{mo}/2 & \Delta_o 
 \end{array} \right)
 	 \left(\begin{array}{c}
		 a_i\\
		 a_m\\ 
		 a_o 
	 \end{array} \right),
\end{equation}
where we have taken the energy of the ground state of the middle ring as the energy origin, and the definitions of the tunnelling rates and energy biases are obtained analogously to Eqs. (\ref{definitionsJD}) with the corresponding localized wavefunctions.
From the diagonalization of this Hamiltonian, one obtains three energy eigenstates $\Psi_{\pm}$ and $\Psi_d$ with eigenvalues $E_{\pm}$ and $E_d$. For $\Delta_o = \Delta_i$, $\Psi_d$, the so-called spatial dark state, involves only the localized ground states of the inner and outer rings,
\begin{equation}
\Psi_d (\Theta)= \cos \Theta \widetilde{\Psi}_i - \sin \Theta \widetilde{\Psi}_o,
\label{darkstate3s}
\end{equation}
with
\begin{equation}
\tan \Theta = J_{im} / J_{mo} .
\label{eq:angle3S}
\end{equation} 

\begin{figure}
\centerline{  \includegraphics[width=0.95\textwidth]{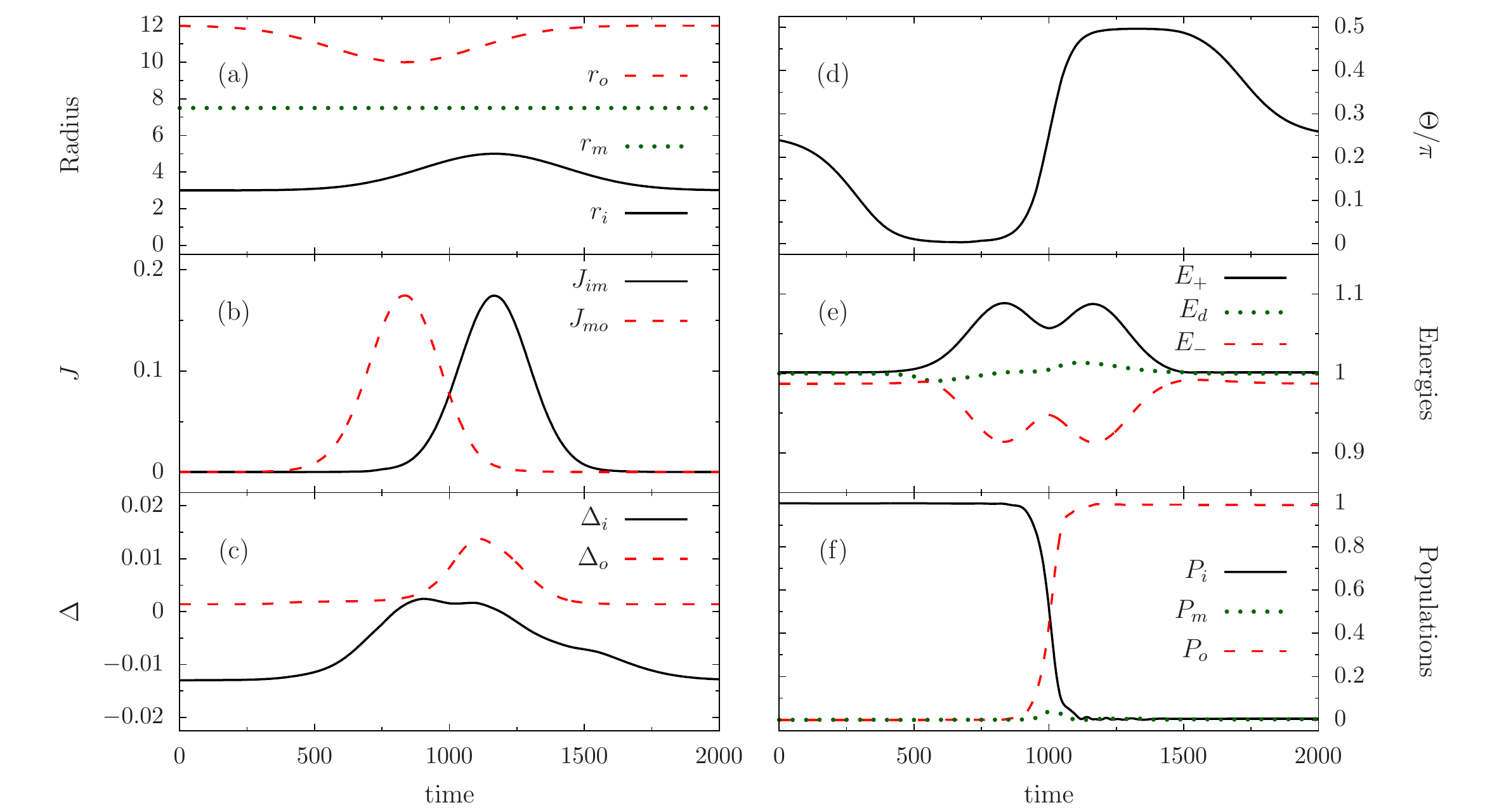}}
	\caption{Numerical simulations using the three-state model. Matter-wave STIRAP process to transport a single atom from the innermost to the outermost potential of three concentric rings, see Fig.~\ref{fig:1POT}(c). As a function of time: (a) radii $r_i$ (black solid line), $r_m$ (green dotted line) and $r_o$ (red dashed line) of the inner, middle and outer rings; (b) tunneling rates $J_{im}$ (black solid line) and $J_{mo}$ (red dashed line); (c) energy bias $\Delta_i$ (black solid line) and $\Delta_o$ (red dashed line) of the inner and outer rings with respect to the middle one; (d) mixing angle $\Theta$; (e)  eigenvalues of the energy eigenstates $\Psi_+$ (black solid line), $\Psi_d$ (green dotted line), and $\Psi_-$ (red dashed line); (f) populations $P_i=|a_i|^2$ (black solid line), $P_m=|a_m|^2$ (green dotted line), and $P_o=|a_o|^2$ (red dashed line) of the inner, middle, and outer rings. All variables are expressed in h.o. units with respect to the harmonic potential $V_h$.}
	\label{fig:5STIRAP_rings}
\end{figure}

\subsection{STIRAP-like protocol}

SAP for a single atom between the innermost and outermost rings can be achieved by adiabatically following $\Psi_d$. Starting with the atom in the inner ring, $\Psi_{3S}(t_0)=\Psi_d(t_0)=\widetilde{\Psi}_i$, and employing a temporal variation of the tunneling rates such that $\Theta$, see (\ref{eq:angle3S}), slowly changes from $0$ to $\pi/2$, one obtains $\Psi_{3S}(t_f)=\Psi_d(t_f)=-\widetilde{\Psi}_o$ at the end of the process. Note that this temporal variation of the mixing angle means to favor first tunneling between the middle and outer rings and later on, and with an appropriate temporal delay, tunneling between the inner and middle rings.

Within the three-state model, we have investigated the transport of a single atom from the localized ground state of the inner ring to the outer one.  Fig.~\ref{fig:5STIRAP_rings}(a) shows the temporal variation of the inner and outer rings radii to achieve the proper variation of the tunneling rates, shown in Fig.~\ref{fig:5STIRAP_rings}(b).
In this case, the time dependences of the radii are described as in the previous cases by Eq. (15b), with $R_0= 3, \delta_R = 2$ for the inner ring and $R_0= 12, \delta_R = -2$ for the outer one. The middle ring has a fixed radius $r_m = 7.5$, and, since no detuning is necessary, all ring trapping frequencies are kept constant and equal to $\omega_i=\omega_m=\omega_o=2$.
The parameter setting chosen implies that the orthonormal localized ground states are not fully resonant during the whole dynamics, see Fig.~\ref{fig:5STIRAP_rings}(c).
However, since $| \Delta_o - \Delta_i |\ll J_{im}, J_{mo}$ in the time window where tunneling is non-negligible, it is possible to follow $\Psi_d$ from $\Theta=0$ to $\Theta=\pi/2$ in this interval, as depicted in Fig.~\ref{fig:5STIRAP_rings}(d).  This adiabatic following results in a high fidelity single atom transport from the inner ring to the outer one, with an almost negligible population in the middle one for the whole dynamics, see Fig.~\ref{fig:5STIRAP_rings}(f).    
\begin{figure}

\centerline{\includegraphics[width=0.8\textwidth]{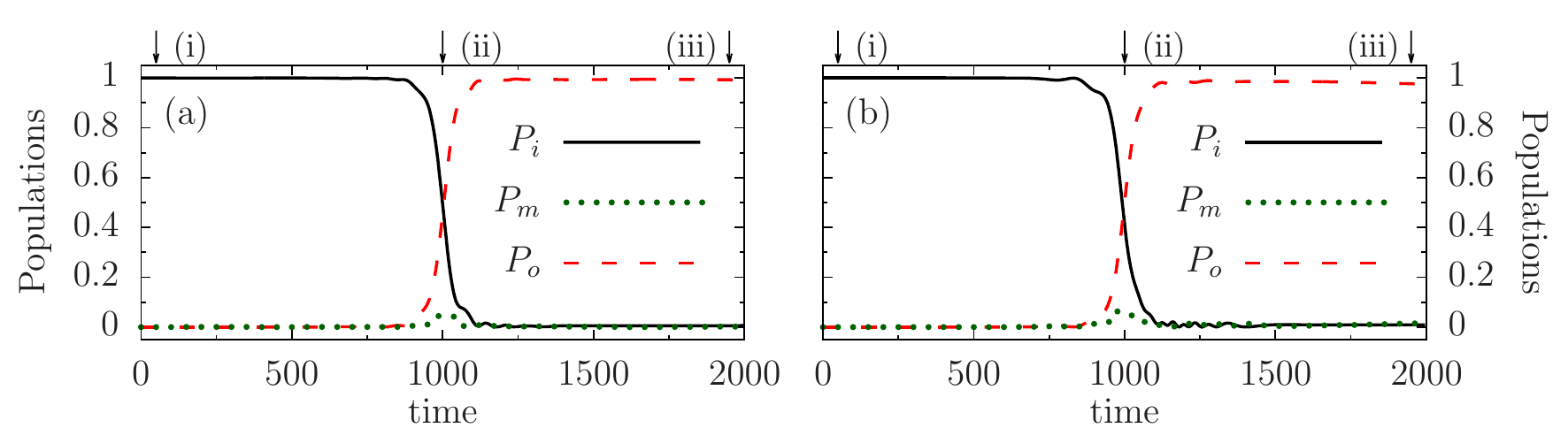} }
\centerline{\includegraphics[width=0.8\textwidth]{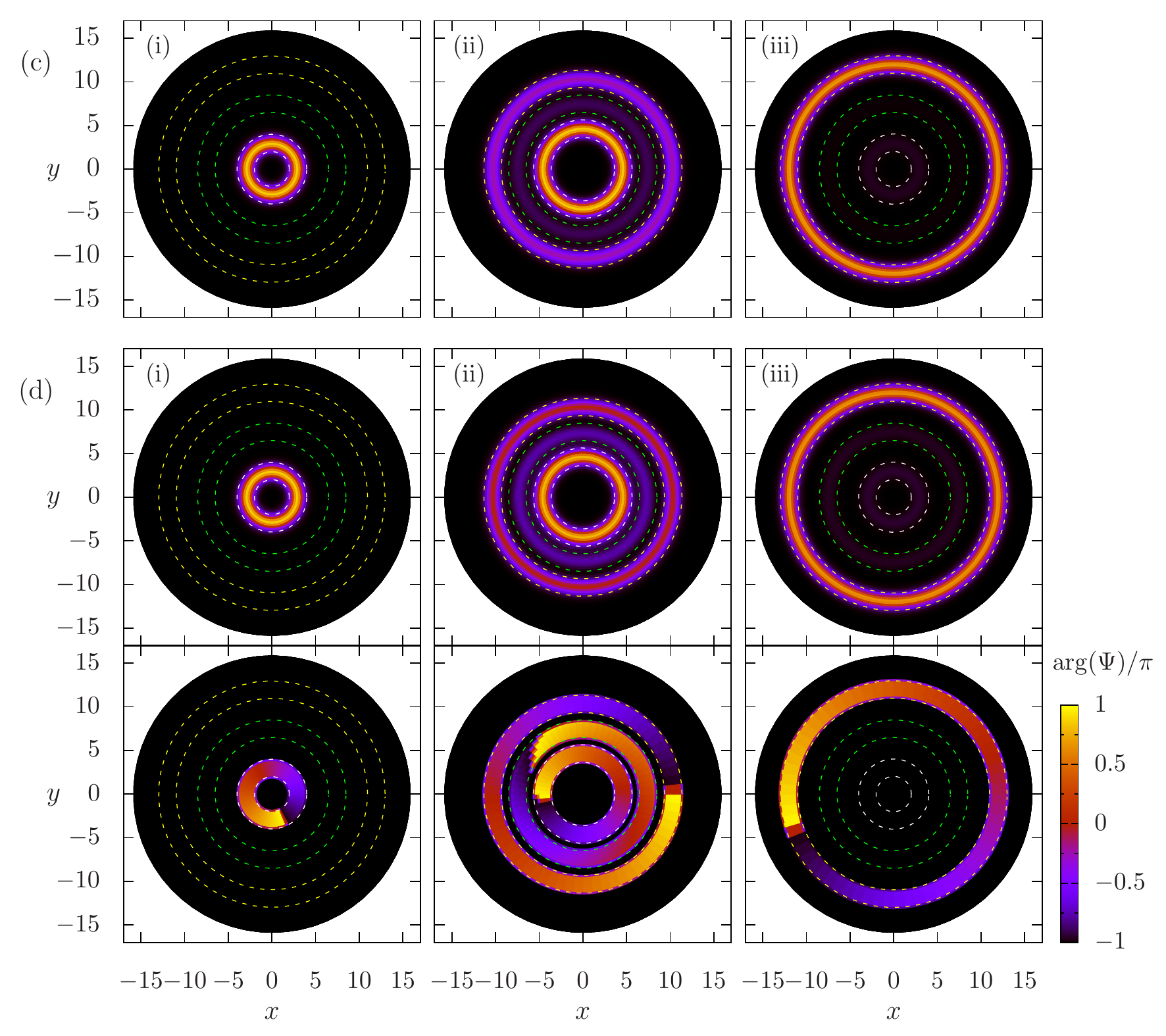} }
	\caption{
	Numerical integration of the 2D Schr\"odinger equation to simulate the matter-wave STIRAP process to transport a single atom from the innermost to the outermost potential of three concentric rings, see Fig.~\ref{fig:1POT}(c).
	Time evolution of the populations of the inner, middle and outer rings when the initial state (obtained from imaginary time evolution) (a) does not posses OAM or (b) carries one unit of OAM.
	(c) Snapshots of the 2D atomic probability density for the three times indicated in (a).
	(d) Snapshots of the 2D atomic probability density (top row) and phase distribution (bottom row) for the three times indicated in (b).
	Parameters are the same as in Fig.~\ref{fig:5STIRAP_rings}.
		}
	\label{fig:6STIRAP_snapshots}
\end{figure}

Fig.~\ref{fig:6STIRAP_snapshots}(a)  shows the temporal evolution of the populations in the inner, middle, and outer rings, respectively, obtained from the numerical integration of the corresponding 2D Schr\"odinger equation for the same parameter values as used for Fig.~\ref{fig:5STIRAP_rings}, and where the initial ground state is obtained via imaginary time evolution. The fidelity of the SAP process is above 99\%.
Fig.~\ref{fig:6STIRAP_snapshots} (c) shows snapshots of the 2D probability density for the three different times indicated in the temporal evolution.
The same protocol applied to a state with OAM is plotted in Figs.~\ref{fig:6STIRAP_snapshots}(b) and (d).
The initial ground state is taken to be the one obtained via imaginary time evolution with an added azimuthal phase $\exp(i \phi)$, and thus with a winding number of one.
The time evolution of the populations from the 2D Schr\"odinger equation in Fig.~\ref{fig:6STIRAP_snapshots}(b) shows a final fidelity above 97\%.
Moreover, Fig.~\ref{fig:6STIRAP_snapshots} (d) shows snapshots for the three different times indicated in the temporal evolution of both the 2D probability density  (top row) and   the corresponding phase distribution (bottom row). Note, comparing the phase distribution in (d,i) with that in (d,iii), that initial and final states have the same OAM winding number. This is because the joint trapping potential consists of three concentric rings and, therefore, cylindrical symmetry is preserved during the whole dynamics. 

\section{Conclusions}
\label{sec:conclusions}

We have investigated SAP processes for the transport of a single-atom from an inner trap to an outer one in three different configurations of cylindrically symmetric concentric potentials. In all cases, the energy eigenstate followed has been transformed from the localized ground state of the inner potential to the localized ground state of the outer ring by manipulating the trapping frequencies and/or the radii of the rings.

Time-averaged adiabatic potentials would be ideal systems to perform these techniques because of their strong confinements on the order of kHz, their possibility of realising trapping lifetimes on the order of up to one minute \cite{rings_4} and their ability to dynamically change the radius of the traps \cite{rings_3}.
Our estimations show that the techniques presented in this paper can be implemented in these systems with a total process time on the order of seconds.
Similar total times can be expected for the painted potentials \cite{rings_5,rings_6}.

We have demonstrated the high efficiency of these SAP processes by means of simple two- and three-state models.
Additionally, we have checked the accuracy of these models by comparing their predictions with the numerical integration of the corresponding 2D Schr\"odinger equation.
The latter has also been used to demonstrate that SAP processes between cylindrically symmetric concentric potentials can be applied to transport states carrying OAM.

\section*{Acknowledgements}	 

J.P., V.A., and J.M. gratefully acknowledge financial support through the Spanish MINECO contract FIS2014-57460-P, and the Catalan Government contract SGR2014-1639.
J.P and J.M. also acknowledge financial support from the FPI Grant No. BES-2012-05344 and the JSPS Research Fellowship S-15025, respectively.
T.B. is grateful to JSPS for partial support from Grant-in-Aid for Scientific Research (Grant No. 26400422).
This work was also supported by the Okinawa Institute of Science and Technology Graduate University.


\section*{References}
\begin{thebibliography}{99}

\bibitem{rings}
E. M. Wright, J. Arlt, and K. Dholakia, Phys. Rev. A \textbf{63}, 013608 (2000);
E. Courtade, O. Houde, J.-F. Clement, P. Verkerk, and D. Hennequin, Phys. Rev. A \textbf{74}, 031403 (2006);
S. E. Olson, M. L. Terraciano, M. Bashkansky, and F. K. Fatemi, Phys. Rev. A \textbf{76}, 061404 (2007);
C. Ryu, M. F. Andersen, P. Clade, V. Natarajan, K. Helmerson, and W. D. Phillips, Phys. Rev. Lett. \textbf{99}, 260401 (2007);
S. Franke-Arnold, J. Leach, M. J. Padgett, V. E. Lembessis, D. Ellinas, A. J. Wright, J. M. Girkin, P. Ohberg, and A. S. Arnold,  Opt. Express \textbf{15}, 8619 (2007);
N. Houston, E. Riis, and A. S. Arnold, J. Phys. B: At. Mol. Opt. Phys. \textbf{41}, 211001 (2008);
A. I. Yakimenko, Yu. M. Bidasyuk, O. O. Prikhodko, S. I. Vilchinskii, E. A. Ostrovskaya, and Yu. S. Kivshar, Phys. Rev. A \textbf{88}, 043637 (2013);
L. Corman, L. Chomaz, T. Bienaime, R. Desbuquois, C. Weintenberg, S. Nascimb\`ene, J. Dalibard, and J. Beugnon, Phys. Rev. Lett. \textbf{113}, 135302 (2014).


\bibitem{rings_2}
G. Birkl, F. B. J. Buchkremer, R. Dumke, and W. Ertmer, Opt. Commun. \textbf{191}, 67 (2001);
A. Turpin, J. Polo, Yu. V. Loiko, J. K\"uber, F. Schmaltz, T. K. Kalkandjiev, V. Ahufinger, G. Birkl, and J. Mompart, Opt. Express \textbf{23}, 1638 (2015).

\bibitem{rings_3}
B. E. Sherlock, M. Gildemeister, E. Owen, E. Nugent, and C. J. Foot, Phys. Rev. A \textbf{83}, 043408 (2011).

\bibitem{rings_4}
I. Lesanovsky, W. von Klitzing, Phys. Rev. Lett. \textbf{99}, 083001 (2007).

\bibitem{rings_5}
S. K. Schnelle, E. D. van Ooijen, M. J. Davis, N. R. Heckenberg, and H. Rubinsztein-Dunlop, Opt. Express \textbf{16}, 1405 (2008).

\bibitem{rings_6}
K. Henderson, C. Ryu, C. MacCormick and M. G. Boshier,
New J. Phys. \textbf{11}, 043030 (2009).


\bibitem{Sagnac}
Y. Japha, O. Arzouan, Y. Avishai, and R. Folman, Phys. Rev. Lett. \textbf{99}, 060402 (2007);
J. L. Helm, S. L. Cornish, and S. A. Gardiner, Phys. Rev. Lett. \textbf{114}, 134101 (2015).

\bibitem{atomtronics}
A. Ruschhaupt and J. G. Muga, Phys. Rev. A \textbf{70}, 061604 (2004);
J. A. Stickney, D. Z. Anderson, and A. A. Zozulya, Phys. Rev. A \textbf{75}, 013608 (2007);
B. T. Seaman, M. Kramer, D. Z. Anderson, and M. J. Holland, Phys. Rev. A \textbf{75}, 023615 (2007);
J. J. Thorn, E. A. Schoene, T. Li, and D. A. Steck, Phys. Rev. Lett. \textbf{100}, 240407 (2008);
R. A. Pepino, J. Cooper, D. Z. Anderson, and M. J. Holland, Phys. Rev. Lett. \textbf{103}, 140405 (2009);
A. Benseny, S. Fern\'andez-Vidal, J. Bagud\`a, R. Corbal\'an, A. Pic\'on, L. Roso, G. Birkl, and J. Mompart,
Phys. Rev. A \textbf{82}, 013604 (2010).

\bibitem{persistentcurrents}
J. Javanainen, S. M. Paik, and S. M. Yoo, Phys. Rev. A \textbf{58}, 580 (1998);
L. Salasnich, A. Parola, and L. Reatto, Phys. Rev. A \textbf{59}, 2990 (1999);
L. Plaja and J. San Rom\'an, Phys. Rev. A \textbf{69}, 063612 (2004);
L. Amico, A. Osterloh, and F. Cataliotti, Phys. Rev. Lett. \textbf{95}, 063201 (2005);
M. F. Andersen, C. Ryu, Pierre Clad\'e, Vasant Natarajan, A. Vaziri, K. Helmerson, and W. D. Phillips, Phys. Rev. Lett. \textbf{97}, 170406 (2006);
A. M. Rey, K. Burnett, I. I. Satija, and C. W. Clark, Phys. Rev. A \textbf{75}, 063616 (2007);
T. Wang, J. Javanainen, S. F. Yelin, Phys. Rev. A \textbf{76}, 011601 (2007);
A. Ramanathan K. C. Wright, S. R. Muniz, M. Zelan, W. T. Hill, C. J. Lobb, K. Helmerson, W. D. Phillips, and G. K. Campbell, Phys. Rev. Lett. \textbf{106}, 130401 (2011).

\bibitem{persistentcurrents_experiment1}
M. F. Andersen, C. Ryu, Pierre Clad\'e, Vasant Natarajan, A. Vaziri, K. Helmerson, and W. D. Phillips, Phys. Rev. Lett. \textbf{97}, 170406 (2006).

\bibitem{persistentcurrents_experiment2}
A. Ramanathan K. C. Wright, S. R. Muniz, M. Zelan, W. T. Hill, C. J. Lobb, K. Helmerson, W. D. Phillips, and G. K. Campbell, Phys. Rev. Lett. \textbf{106}, 130401 (2011).

\bibitem{new_atomtronics}
K. C. Wright, R. B. Blakestad, C. J. Lobb, W. D. Phillips, and G. K. Campbell, Phys. Rev. Lett. \textbf{110}, 025302 (2013);
C. Ryu, P. W. Blackburn, A. A. Blinova, and M. G. Boshier, Phys. Rev. Lett. \textbf{111}, 205301 (2013);
F. Jendrzejewski, S. Eckel, N. Murray, C. Lanier, M. Edwards, C. J. Lobb, and G. K. Campbell, Phys. Rev. Lett. \textbf{113}, 045305 (2014);
D. Aghamalyan, M. Cominotti, M. Rizzi, D. Rossini, F. Hekking, A. Minguzzi, L.C. Kwek and L. Amico, New J. Phys., \textbf{17}, (2015);
Davit Aghamalyan, Luigi Amico, and L. C. Kwek, Phys. Rev. A \textbf{88}, 063627 (2013);
Yi-Hsieh Wang, A. Kumar, F. Jendrzejewski, Ryan M. Wilson, Mark Edwards, S. Eckel, G. K. Campbell, Charles W. Clark, arXiv:1510.02968 [cond-mat.quant-gas];
R. Mathew, A. Kumar, S. Eckel, F. Jendrzejewski, G. K. Campbell, Mark Edwards, and E. Tiesinga, Phys. Rev. A \textbf{92}, 033602 (2015).

\bibitem{rings_1}
S. Eckel, J. G. Lee, F. Jendrzejewski, N. Murray, C. W. Clark, C. J. Lobb, W. D. Phillips, M. Edwards, and G. K. Campbell, Nature \textbf{506}, 200 (2014).


\bibitem{Brand_2009}
J. Brand, T. J. Haigh, and U. Z\"ulicke, Phys. Rev. A \textbf{80}, 011602(R) (2009). 


\bibitem{Zhang_2012}
X. F. Zhang, R. F. Dong, T. Liu, W. M. Liu, and S. G. Zhang, Phys. Rev. A \textbf{86}, 063628 (2012).

\bibitem{Malet_2010}
F. Malet, G. M. Kavoulakis, and S. M. Reimann, Phys. Rev. A \textbf{81}, 013630 (2010).

\bibitem{review_SAP}
R. Menchon-Enrich, A. Benseny, V. Ahufinger, A. D. Greentree, Th. Busch, and J. Mompart, submitted to Reports on Progress in Physics (2015).

\bibitem{eckert1}
K. Eckert, M. Lewenstein, R. Corbal\'an, G. Birkl, W. Ertmer, and J. Mompart, Phys. Rev. A \textbf{70}, 023606 (2004).

\bibitem{eckert2}
K. Eckert, J. Mompart, R. Corbal\'an, M. Lewenstein, and G. Birkl, Optics Comm. \textbf{264}, 264 (2006).

\bibitem{singleatom_1DSAP}
T. Opatrn\'y and K. K. Das, Phys. Rev. A \textbf{79}, 012113 (2009);
S. McEndoo, S. Croke, J. Brophy, and Th. Busch, Phys. Rev. A \textbf{81}, 043640 (2010);
T. Morgan, B. O'Sullivan, and Th. Busch, Phys. Rev. A \textbf{83}, 053620 (2011);
A. Benseny, Joan Bagud\`a, X. Oriols, and J. Mompart, Phys. Rev. A \textbf{85}, 053619 (2012).

\bibitem{loiko}
Yu. Loiko, V. Ahufinger, R. Corbal\'an, G. Birkl, and J. Mompart
Phys. Rev. A \textbf{83}, 033629 (2011).

\bibitem{morgan}
T. Morgan, L. J. O'Riordan, N. Crowley, B. O'Sullivan, and Th. Busch, Phys. Rev. A \textbf{88}, 053618 (2013).

\bibitem{electrons_1DSAP}
A. D. Greentree, J. H. Cole, A. R. Hamilton, and L. C. L. Hollenberg, Phys. Rev. B \textbf{70}, 235317 (2004).

\bibitem{BEC_1DSAP}
E. M. Graefe, H. J. Korsch, and D. Witthaut. Phys. Rev. A \textbf{73}, 013617 (2006);
M. Rab, J. H. Cole, N. G. Parker, A. D. Greentree, L. C. L. Hollenberg, and A. M. Martin, Phys. Rev. A \textbf{77}, 061602 (2008);
V. O. Nesterenko, A. N. Novikov, F. F. de Souza Cruz, and E. L. Lapolli, Laser Physics \textbf{19}, 616 (2009);
M. Rab, A. L. C. Hayward, J. H. Cole, A. D. Greentree, and A. M. Martin, Phys. Rev. A \textbf{86}, 063605 (2012).

\bibitem{interferometry_2DSAP}
R. Menchon-Enrich, S. McEndoo, Th. Busch, V. Ahufinger, and J. Mompart, Phys. Rev. A \textbf{89}, 053611 (2014).

\bibitem{OAM_2DSAP}
R. Menchon-Enrich, S. McEndoo, J. Mompart, V. Ahufinger, and Th. Busch, Phys. Rev. A \textbf{89}, 013626 (2014).

\bibitem{light_SAP}
S. Longhi, G. Della Valle, M. Ornigotti, and P. Laporta, Phys. Rev. B \textbf{76}, 201101 (2007);
Y. Lahini, F. Pozzi, M. Sorel, R. Morandotti, D. N. Christodoulides, and Y. Silberberg, Phys. Rev. Lett. \textbf{101}, 193901 (2008); R. Menchon-Enrich, A. Llobera, V. J. Cadarso, J. Mompart, and V. Ahufinger, IEEE Photonics Technology Lett. \textbf{24}, 536 (2012); R. Menchon-Enrich, A. Llobera, J. Vila-Planas, V. J. Cadarso, J. Mompart, and V. Ahufinger, Light: Science \& Applications \textbf{2}, e90 (2013).

\bibitem{sound_SAP}
R. Menchon-Enrich, J. Mompart, and V. Ahufinger, Phys. Rev. B \textbf{89}, 094304 (2014).

\bibitem{RAP}
N. V. Vitanov, T. Halfmann, B. W Shore, and K. Bergmann, Annual Review of Physical Chemistry \textbf{52}, 763 (2001).

\bibitem{STIRAP}
U. Gaubatz, P. Rudecki, S. Schiemann, and K. Bergmann, J. Chem. Phys. \textbf{92}, 5363 (1990);  
K. Bergmann, H. Theuer, and B. W. Shore, Rev. Mod. Phys. \textbf{70}, 1003 (1998).






\end{thebibliography}
\end{document}